\documentclass{aa}
\usepackage{times}
\usepackage{graphicx}

\usepackage{amsmath}
\usepackage{color}
\usepackage{graphics}
\usepackage{natbib}
\usepackage{txfonts}
\usepackage{subfigure}

\bibpunct{(}{)}{;}{a}{}{,}

\newcommand {\be}{\begin{equation}} 
\newcommand{\ee}{\end{equation}}    
\def\nabp{\nabla_{\perp}}
\newcommand{\sss}{\scriptscriptstyle}
\def\ddt{\frac{\partial}{\partial t}}

\def\nabp{\nabla_{\perp}}
\begin{document}
%
%
\title{Comment on "Drift instabilities in the solar corona within the multi-fluid description" [Astron.\ Astrophys.\ {\bf 481}, 853 (2008)]}

\author{J. Vranjes\inst{1,2}
     \and
S. Poedts\inst{1}
       }


\institute{Center  for  Plasma Astrophysics, and Leuven Mathematical Modeling and Computational Science Center
 (LMCC), Celestijnenlaan 200 B, 3001
Leuven, Belgium
 \and
Facult\'{e} des Sciences Appliqu\'{e}es, avenue F.D. Roosevelt 50,
 1050 Bruxelles, Belgium\\
           \email{Jovo.Vranjes@wis.kuleuven.be;
Stefaan.Poedts@wis.kuleuven.be}
           }
   \date{Received \ Accepted}

\abstract
{Mecheri $\&$ Marsch (2008) concluded that drift currents caused by density gradients can serve as an energy source for a plasma instability, in particular for the excitation of ion-cyclotron waves in the solar corona. }
{It is pointed out that these authors overlooked some fundamental properties of the drift motion in inhomogeneous plasmas.}
{The calculation is repeated, taking into account the missing terms. }
{It is shown that the diamagnetic drift, which is essential for the new physical phenomena obtained by Mecheri $\&$ Marsch (2008),  can not contribute to the flux in the continuity equation.
Moreover, the part of the ion polarization drift contribution to the ion cancels out exactly with the contribution of the part of the stress tensor drift to the same flux.}
{The ion-cyclotron waves in the solar corona can thus not be excited in the way suggested by Mecheri $\&$ Marsch (2008).}

\titlerunning{Comment}
\authorrunning{Vranjes \& Poedts}


\maketitle
%

Recently, Mecheri $\&$ Marsch (2008) discussed the drift instability in an application to the solar corona. The conclusion was made that drift currents caused by density gradients can serve as an energy source for plasma instability, in particular for the excitation of ion-cyclotron waves. The multi-fluid theory was used in a collision-less limit. The equilibrium diamagnetic drift currents are essential for all the new physical phenomena presented in the work.

However, the authors have overlooked some fundamental properties of the drift motion in inhomogeneous plasmas. We stress that:
\begin{description}
  \item[(i)] The diamagnetic drift is a fluid effect and not a particle drift, therefore it can not contribute to the flux in the continuity equation.
  \item[(ii)] There exists a well-known cancelation of the part of the ion polarization drift contribution to the ion flux from one side, and the contribution of the part of the stress tensor drift to the same flux, from the other.
\end{description}

These two facts describe some fundamental properties of drift motions in inhomogeneous plasmas \cite{wei}, \cite{vr2}, \cite{vr1}. A more detailed picture is obtained from the following consideration. Assume a plasma embedded in a homogeneous magnetic field in the  $z$-direction, $B_0 \vec e_z$, and with an equilibrium plasma density that has a gradient in the direction perpendicular to the magnetic field vector.  The effect of an additional  inhomogeneity of the magnetic field, as discussed  by Mecheri $\&$ Marsch,  will be commented below.
Using the momentum equations for ions and massless electrons
\[
 m_in_i\left[\frac{\partial \vec v_i}{\partial t} + (\vec
v_i\cdot\nabla) \vec v_i\right] = e n_i \left(-\nabla \phi  - \frac{\partial
A_z}{\partial t} \vec e_z + \vec v_i\times \vec B\right)
\]
\be
- \kappa T_i\nabla n_i - \nabla\cdot \Pi_i, \label{1} \ee
\be
0 = -e n_i \left(-\nabla \phi  - \frac{\partial
A_z}{\partial t} \vec e_z + \vec v_e\times \vec B\right)
 - \kappa T_e \nabla n_e - \nabla\cdot \Pi_e,
 \label{2} \ee
the total perpendicular velocities for the two species, without approximations, can be written as:
\[
v_{i\bot}=\frac{1}{B_0} \vec e_z \times \nabp \phi
 + \frac{v_{{\sss T}
i}^2}{\Omega_i} \vec e_z \times \frac{\nabp n_i}{n_i} + \vec
e_z \times \frac{\nabp\cdot \Pi_i}{m_i n_i \Omega_i}
\]
\be
    + \frac{1}{\Omega_i}
\left(\ddt + \vec v_i  \cdot \nabla\right) \vec e_z \times \vec v_{i
\bot},
\label{3} \ee
\be \vec v_{e\bot }= \frac{1}{B_0} \vec e_z\times \nabp \phi  - \frac{\kappa
T_e}{e B_0} \vec e_z \times \frac{\nabla_\bot n_e}{n_e}. \label{4} \ee
Here,
$\vec B=\nabla\times \vec A=- \vec e_z\times \nabla_\bot A_{z}$, $\vec E=-\nabla \phi - \partial \vec A_{z}/\partial t$, and the electron stress tensor terms are neglected.
The diamagnetic drift $\vec v_{dj}$  corresponds to the second term on the right-hand sides in Eqs.~(\ref{3}) and (\ref{4}).

Related to remark (i) above, it is seen that
\be
\nabla\cdot (n_j\vec v_{dj})\equiv 0, \label{5}
\ee
as long as the magnetic field is homogeneous, describing a well known fundamental property, see e.g.,  Weiland (2000). Hence, this term does not contribute to the flux in the continuity equations. However, this term is essential in the commented work, where it gives rise to the term  $(\vec k \cdot \vec v_{dj}) n_{j1}$ in the continuity equation [Eq.~(10)], and it is exactly this term that provides the source of the current-driven instability discussed in the paper.
We note that, strictly speaking, the magnetic field in the work of Mecheri $\&$ Marsch is indeed inhomogeneous. Yet, in this particular and essential term $(\vec k \cdot \vec v_{dj}) n_{j1}$, the magnetic field inhomogeneity is explicitly omitted as a higher order correction [cf., Eq.~(13)  and the related  text], and consequently  the condition (\ref{5}) given above holds within the same approximation.

Equally important is the issue (ii) regarding the cancelation of terms related to the stress tensor drift and the polarization drift.
This appears as follows.

First we note that  the polarization drift $\vec v_p$ is the last term in Eq.~(\ref{3}), and its convective derivative
 \be
 (\vec v_i\cdot\nabla)\vec e_z\times \vec v_{i\bot},
 \label{6}
 \ee
is the one that takes part in the cancelation.
The procedure is described in detail in Weiland (2000) and  Vranjes $\&$ Poedts (2006a).
Within the same approximations as in the commented paper (small gradients of the equilibrium quantities),  the last $\vec v_{i\bot}$ in Eq.~(\ref{6})  contains only the leading
order perturbed drifts  from Eq.~(\ref{3}).
On the other hand, the first $\vec v_i$ in Eq.~(\ref{6}) can only be  the equilibrium ion diamagnetic drift. This is then to be used in the term $\nabla\cdot(n_i \vec v_p)$ in the continuity equation.

On the other hand, the stress tensor part [the third term on the right-hand side in Eq.~(\ref{3})] yields
\be
\nabp\cdot(n \vec v_\pi) =-\rho_i^2 \nabp n_{i0} \cdot \nabp^2 \vec v_{i\bot} -
n_{i0} \rho_i^2 \nabp^2 \nabp \cdot \vec v_{i\bot}. \label{7} \ee
Within the second-order approximation limit, the first term on the right-hand side in this expression cancels out exactly with the  above discussed convective derivative in the polarization
drift, see Weiland (2000), Vranjes $\&$ Poedts (2006a). Hence, the effect of the  equilibrium diamagnetic drift (that is crucial for Mecheri $\&$ Marsch work)  vanishes here as well.

We observe  that the stress tensor contribution is nowhere  included in the work of Mecheri $\&$ Marsch. This implies that their  equations contain extra terms originating from the polarization drift, and which in reality cancel out exactly with the stress tensor part, as described above. These terms are in fact  explicitly seen in Eq.~(11) of the commented paper, containing  the critical term $\vec k \cdot \vec v_{dj}$ in the  polarization drift term, equivalent to Eq.~(\ref{6}).

To conclude, we have shown that the results of Mecheri $\&$ Marsch are an artifact of some basic  errors in the starting set of equations. Their equations indeed contain the terms which determine all the  results obtained in the commented paper, however  those terms cancel out exactly due to Eq. (\ref{5}) and due to the stress tensor effect. This cancelation is valid in general and independent on the physical system, and it must  be taken into account in a proper model.

\begin{acknowledgements}

These results  are  obtained in the framework of the
projects G.0304.07 (FWO-Vlaanderen), C~90205 (Prodex),  GOA/2004/01
(K.U.Leuven),  and the Interuniversity Attraction Poles Programme -
 Belgian State - Belgian Science Policy.

\end{acknowledgements}

\end{document}